\documentclass[10pt,twocolumn]{article}

\usepackage[margin=0.75in]{geometry}

\usepackage{amsmath,amssymb,bm}
\usepackage{graphicx}
\usepackage{booktabs}

\usepackage[numbers,sort&compress]{natbib}

\usepackage[hidelinks]{hyperref}

\usepackage{epstopdf}

\title{Decomposing the Influence of Physical Acoustic Modeling on Neural Personal Sound Zone Rendering: An Ablation Study\\
}
\author{
Hao Jiang$^{1}$ \quad Edgar Choueiri$^{1}$\\
\small $^{1}$3D Audio and Applied Acoustics Laboratory, Princeton University, Princeton, NJ, USA\\
\small \texttt{hj3737@princeton.edu, choueiri@princeton.edu}
}
\date{}

\begin{document}
\maketitle
\begin{abstract}%
Deep learning-based Personal Sound Zones (PSZs) rely on simulated acoustic transfer functions (ATFs) for training, yet idealized point-source models exhibit large sim-to-real gaps. While physically informed components improve generalization, individual contributions remain unclear. This paper presents a controlled ablation study on a head-pose-conditioned binaural PSZ renderer using the Binaural Spatial Audio Neural Network (BSANN). We progressively enrich simulated ATFs with three components: (i) anechoically measured frequency responses of the particular loudspeakers(FR), (ii) analytic circular-piston directivity (DIR), and (iii) rigid-sphere head-related transfer functions (RS-HRTF). Four configurations are evaluated via in-situ measurements with two dummy heads. Performance metrics include inter-zone isolation (IZI), inter-program interference (IPI), and crosstalk cancellation (XTC) over 100--20{,}000~Hz. Results show FR provides spectral calibration, yielding modest XTC improvements and reduced inter-listener IPI imbalance. DIR delivers the most consistent sound-zone separation gains (10.05~dB average IZI/IPI). RS-HRTF dominates binaural separation, boosting XTC by +2.38/+2.89~dB (average 4.51 to 7.91~dB), primarily above 2~kHz, while introducing mild listener-dependent IZI/IPI shifts. These findings guide prioritization of measurements and models when constructing training ATFs under limited budgets.
\end{abstract}

\section{Introduction} \label{sec:intro}

Personal Sound Zones (PSZs) aim to deliver independent audio programs to multiple listeners within a shared acoustic space without headphones by shaping the sound field with a loudspeaker array~\cite{druyvesteyn1997personal,betlehem2015personal}. This capability enables private or personalized listening experiences in domestic and automotive environments, among other shared-space scenarios~\cite{cheer2013car,vindrola2021car}. Classical PSZ reproduction is typically posed as an optimization problem based on measured acoustic transfer functions (ATFs), with representative approaches including acoustic contrast control (ACC)~\cite{choi2002brightzone,coleman2014acousticcontrast} and pressure matching (PM)~\cite{poletti2008multizone}. Variants and extensions continue to refine objective formulations and perceptual trade-offs, such as time-domain optimization and variable span trade-off filtering~\cite{galvez2015timedomain,lee2020signaladaptive,brunnstrom2022vast,abe2023amplitude}.

A persistent challenge for PSZ systems is robustness to mismatch between the measured ``plant'' ATFs used for design and the transfer functions encountered during operation. Such mismatch may arise from listener movement, repositioned loudspeakers, changes in room conditions, and measurement noise, which can substantially reduce isolation and increase interference~\cite{elliott2012robustness,moller2019tfnoise}. Accordingly, a range of robustification strategies has been investigated, including regularization and reduced-measurement schemes aided by acoustic modeling~\cite{elliott2012robustness,zhu2017robustacc}. For dynamic settings, adaptive and position-dependent approaches have also been explored to accommodate moving listeners and time-varying conditions~\cite{vindrola2021car,galvez2019dynamic,lindfors2022equalization}.

Recently, deep learning-based methods have emerged as a complementary paradigm for PSZ rendering and filter design. State-conditioned neural renderers, such as the Spatial Audio Neural Network (SANN)~\cite{qiao2025sann}, amortize filter design into learning, thereby reducing run-time computational burden and potentially improving robustness when trained over sufficiently diverse acoustic conditions. Nevertheless, the effectiveness of such data-driven methods hinges on the quality and coverage of the transfer functions used to construct training and evaluation data. For head-tracked and binaural rendering in particular, acquiring dense, listener-dependent measured ATF datasets is often prohibitively time-consuming, motivating reliance on simulated ATFs. However, simulations based on idealized point-source assumptions frequently transfer poorly to real deployments, leading to a pronounced sim-to-real gap.

This gap is largely attributable to physical behaviors that idealized point-source models neglect, including device-specific spectral coloration, frequency-dependent loudspeaker directivity, and head-related binaural propagation effects that determine interaural cues. To narrow this gap in a practical way, our recent work on the Binaural Spatial Audio Neural Network (BSANN)~\cite{Jiang2026BSANN} has adopted physically informed ATFs that enrich simulations with transferable components that can be measured or modeled offline, such as anechoically measured loudspeaker responses, analytic piston directivity, and rigid-sphere head-related transfer function (RS-HRTF)~\cite{choueiri2018binaural,Stepanishen1971Transient,duda1998spherical}. These choices are attractive because they avoid the need for exhaustive in-room measurements while capturing key non-ideal loudspeaker and binaural effects that dominate sim-to-real mismatch. However, these components are typically introduced en bloc, leaving the incremental benefit of each modeling layer unclear; for example, whether improvements in crosstalk cancellation versus sound-zone separation are primarily driven by transducer characteristics or by head scattering effects.

This paper addresses this question through a controlled stage-wise ablation study that quantifies how progressively adding physically informed acoustic components affects neural PSZ rendering. Adopting the BSANN architecture~\cite{Jiang2026BSANN} as the representative framework, we employ a progressive enrichment protocol that starts from a baseline point-source simulation and sequentially incorporates: (i) anechoically measured frequency responses of the particular loudspeakers (FR), (ii) analytic piston directivity modeling (DIR)~\cite{Stepanishen1971Transient,morse1968theoretical}, and (iii) rigid-sphere head-related transfer function (RS-HRTF) modeling~\cite{duda1998spherical,kuhn1977spherical}. This yields four configurations: baseline, baseline+FR, baseline+FR+DIR, and baseline+FR+DIR+RS-HRTF. Across all configurations, we keep the neural architecture, optimization settings, data split, and training protocol fixed, varying only the ATF-generation pipeline. While this stage-wise design does not fully decouple component effects from ordering, it reflects a physically motivated propagation chain and enables attribution of performance changes to the incremental inclusion of each modeling layer under this ordering. Performance is quantified using inter-zone isolation (IZI), inter-program interference (IPI) defined in~\cite{qiao2022isolation}, and crosstalk cancellation (XTC), reported both broadband and over representative frequency ranges.

The contributions of this work are threefold: 
\begin{enumerate}
\item a controlled stage-wise ablation protocol that quantifies the impact of progressively enriching simulated ATFs with physically informed acoustic components in neural PSZ training; 
\item quantitative analysis characterizing the incremental benefits of spectral correction (FR), directivity modeling (DIR), and rigid-sphere HRTF modeling (RS-HRTF) on sound-zone separation and XTC, including broadband and frequency-range analyses; 
\item practical guidance on prioritizing physical measurements and models when constructing training ATFs under limited measurement and implementation budgets. \end{enumerate}

\section{Method}
\label{sec:method}

This section details the experimental framework used to isolate the contributions of physically informed acoustic components. We summarize the fixed BSANN renderer at a high level (with full architectural details deferred to~\cite{Jiang2026BSANN}), describe the physically informed ATF construction pipeline and the cumulative ablation protocol, and define the evaluation metrics.

\subsection{Neural PSZ Rendering Framework}
\label{sec:method_renderer}

We employ the Binaural Spatial Audio Neural Network (BSANN)~\cite{Jiang2026BSANN} as the fixed rendering framework for all experiments.
Conditioned on the listener's instantaneous head pose, the network predicts a bank of control filters to achieve dynamic binaural reproduction.
Let $L$ denote the number of loudspeakers and $K$ the number of listeners (programs), where each program $k$ consists of stereo source channels $s_{k,c}[n]$ for $c\in\{L,R\}$. In this work, each listener $k$ is associated with one dedicated program $k$.
For a given state, the renderer generates a filter bank
\begin{equation}
\mathbf{w}=\left\{w_{\ell,k,c}[n]\right\}_{\substack{\ell=1\dots L \\ k=1\dots K \\ c\in\{L,R\}}}.
\end{equation}
The loudspeaker driving signal for unit $\ell$ is formed as a superposition of all convolved source channels:
\begin{equation}
x_\ell[n] = \sum_{k=1}^{K} \sum_{c\in\{L,R\}} (s_{k,c} * w_{\ell,k,c})[n],
\label{eq:drive}
\end{equation}
where $*$ denotes convolution.

BSANN is trained end-to-end using the same objective and protocol as in~\cite{Jiang2026BSANN}, with a composite loss comprising bright-zone accuracy and dark-zone suppression terms, together with gain-limiting and time-domain compactness regularizers. Crucially, to ensure a fair ablation study, the renderer architecture, loss weights, optimization hyperparameters, and training/validation data splits are held constant across all experimental configurations. The only variable is the set of ATFs used to simulate the acoustic environment during training.

\subsection{Physically Informed ATF Modeling}
\label{sec:method_atf}

To synthesize training ATFs, we follow the physically informed ATF construction pipeline in Sec.~II-B of~\cite{Jiang2026BSANN}, in which simulated point-source room impulse responses are decomposed into direct and reflected components. For each listener--ear--control-point--loudspeaker tuple $(k,e,m,\ell)$ (listener $k$, ear $e\in\{L,R\}$, control point $m$, loudspeaker $\ell$), gpuRIR~\cite{DiazGuerra2021gpuRIR} is used to generate a point-source RIR that is separated into a direct component $h^{\mathrm{dir}}_{k,e,m,\ell}[n]$ and a reflected component $h^{\mathrm{refl}}_{k,e,m,\ell}[n]$. Their frequency-domain transfer functions are
\begin{equation}
\begin{split}
H^{\mathrm{dir}}_{k,e,m,\ell}(\omega) &= \mathrm{FFT}\{h^{\mathrm{dir}}_{k,e,m,\ell}[n]\}, \\
H^{\mathrm{refl}}_{k,e,m,\ell}(\omega) &= \mathrm{FFT}\{h^{\mathrm{refl}}_{k,e,m,\ell}[n]\}.
\end{split}
\end{equation}
We then progressively augment the direct and reflected components using three physically informed components: (i) anechoically measured frequency responses of the particular loudspeakers (FR), (ii) analytic piston directivity (DIR), and (iii) rigid-sphere HRTFs (RS-HRTF).

\paragraph*{(i) Anechoic loudspeaker frequency responses of the particular loudspeakers (FR).}
To capture loudspeaker-dependent spectral coloration, we incorporate the anechoically measured complex frequency response $A_\ell(\omega)$ of each loudspeaker unit. The measured response is applied to both the direct and reflected paths.

\paragraph*{(ii) Analytic piston directivity (DIR).}
To model frequency-dependent directivity, we employ the analytic circular-piston model in a rigid (infinite) baffle~\cite{Stepanishen1971Transient,morse1968theoretical}. Let $\theta_{k,e,m,\ell}$ denote the off-axis angle between the loudspeaker acoustic axis and the direction from loudspeaker $\ell$ to the listener-$k$ control point $m$ around ear $e$. The far-field amplitude directivity factor is given by
\begin{equation}
D_\ell(\omega,\theta_{k,e,m,\ell})
=
\frac{2 J_1\!\left(k_0 a_\ell \sin\theta_{k,e,m,\ell}\right)}
{k_0 a_\ell \sin\theta_{k,e,m,\ell}},\
k_0=\frac{\omega}{c},
\label{eq:piston_dir}
\end{equation}
where $J_1(\cdot)$ is the first-order Bessel function of the first kind, $a_\ell$ is the effective piston radius, and $c$ is the speed of sound. The on-axis value is defined by continuity as $D_\ell(\omega,0)=1$. The directivity model is applied only to the direct component, while the reflected component contributes nondirectional late energy.

\paragraph*{(iii) Rigid-sphere HRTFs (RS-HRTF).}
To capture head-induced scattering and shadowing effects, the ear-adjacent control points are modeled using a normalized rigid-sphere scattering formulation~\cite{duda1998spherical,kuhn1977spherical}. We model the head as a rigid sphere of radius $R_k$. Let $\mathbf{r}_{k,\ell}$ denote the vector from listener-$k$ head center to loudspeaker $\ell$, and let $\mathbf{r}_{k,e,m}$ denote the position of control point $m$ around ear $e$ relative to the same head center. We define $r_{k,\ell}=\|\mathbf{r}_{k,\ell}\|$, $r_{k,e,m}=\|\mathbf{r}_{k,e,m}\|$, and the angular separation $\gamma_{k,e,m,\ell}=\angle(\mathbf{r}_{k,\ell},\mathbf{r}_{k,e,m})$.

For acoustic wavenumber $k_0=\omega/c$, the scattering kernel is defined as
\begin{equation}
\label{eq:S_n}
\begin{aligned}
S_n(k_0)
=\;& j_n(k_0 r_<)\,h^{(1)}_n(k_0 r_>) \\
&- \alpha_n(k_0 R_k)\,
   h^{(1)}_n(k_0 r_{k,\ell})\,
   h^{(1)}_n(k_0 r_{k,e,m}),
\end{aligned}
\end{equation}
where $r_< = \min(r_{k,\ell},r_{k,e,m})$, $r_> = \max(r_{k,\ell},r_{k,e,m})$, $j_n(\cdot)$ and $h^{(1)}_n(\cdot)$ are spherical Bessel and Hankel functions, and $\alpha_n(\cdot)$ denotes the rigid-sphere scattering coefficient.

The normalized rigid-sphere HRTF from loudspeaker $\ell$ to control point $m$ around ear $e$ (for listener $k$) is given by
\begin{equation}
\label{eq:rigidsphere_hrtf}
\begin{split}
H^{\mathrm{HRTF}}_{k,e,m,\ell}(\omega) &= \frac{\mathrm{i} k_0}{G_{\mathrm{ff},k,e,m,\ell}(\omega)} \\
&\quad \cdot \sum_{n=0}^{N_{\mathrm{max}}}(2n+1)\, S_n(k_0)\, P_n(\cos\gamma_{k,e,m,\ell}),
\end{split}
\end{equation}
where $P_n(\cdot)$ is the Legendre polynomial, $N_{\mathrm{max}}$ is the truncation order, and
\begin{equation}
G_{\mathrm{ff},k,e,m,\ell}(\omega)
=
\frac{
e^{-\mathrm{i}k_0\|\mathbf{r}_{k,\ell}-\mathbf{r}_{k,e,m}\|}
}{
\|\mathbf{r}_{k,\ell}-\mathbf{r}_{k,e,m}\|
}
\end{equation}
is the free-field Green's function used for normalization. This term is applied exclusively to the direct path component.

\subsection{Cumulative Ablation Protocol}
\label{sec:method_ablation}

We adopt a cumulative ablation strategy that mirrors a physically motivated propagation chain: source transducer response (FR) $\to$ source radiation (DIR) $\to$ head scattering (RS-HRTF). We define four configurations (C0--C3), summarized in Table~\ref{tab:ablation}. Starting from the baseline point-source simulation (C0), we sequentially enable the anechoically measured frequency response (FR), the analytic directivity (DIR), and the rigid-sphere head scattering term (RS-HRTF).

For each configuration, the training ATFs are synthesized by augmenting the point-source direct and reflected components described in Sec.~\ref{sec:method_atf}. 
For notational brevity, we omit the frequency dependence $(\omega)$ and path indices $\{k,e,m,\ell\}$ in the following formulations. 
Let $\theta$ and $\Omega$ denote the specific loudspeaker off-axis angle and head incidence angle, respectively. 
The cumulative configurations are defined as:
\begin{equation}
\label{eq:ablation_assembly}
\begin{aligned}
H^{\mathrm{C0}}
&= H^{\mathrm{dir}} + H^{\mathrm{refl}}, \\[1ex]
H^{\mathrm{C1}}
&= A \Bigl[ H^{\mathrm{dir}} + H^{\mathrm{refl}} \Bigr], \\[1ex]
H^{\mathrm{C2}}
&= A \Bigl[ D(\theta) H^{\mathrm{dir}} + H^{\mathrm{refl}} \Bigr], \\[1ex]
H^{\mathrm{C3}}
&= A \Bigl[ D(\theta) H^{\mathrm{HRTF}}(\Omega) H^{\mathrm{dir}} + H^{\mathrm{refl}} \Bigr].
\end{aligned}
\end{equation}
Equivalently, the baseline C0 is recovered from C3 by setting $A=1$, $D(\theta)=1$, and $H^{\mathrm{HRTF}}(\Omega)=1$. Consistent with Sec.~\ref{sec:method_atf}, the spectral response $A$ is applied to both direct and reflected paths, whereas the spatial terms $D$ and $H^{\mathrm{HRTF}}$ modulate the direct component only.

\begin{table}[t]
\tabcolsep8pt
\caption{Cumulative ablation configurations. Each stage builds upon the previous one by enabling an additional physically informed component.}
\label{tab:ablation}
{%
\begin{tabular}{@{}lccc@{}}\toprule
ID & FR ($A_\ell$) & DIR ($D_\ell$) & RS-HRTF ($H^{\mathrm{HRTF}}$) \\\midrule
C0 & No  & No  & No  \\
C1 & Yes & No  & No  \\
C2 & Yes & Yes & No  \\
C3 & Yes & Yes & Yes \\\bottomrule
\end{tabular}}
\end{table}

\subsection{Metrics}
\label{sec:method_metrics}

We quantify sound-zone separation and binaural performance using inter-zone isolation (IZI), inter-program interference (IPI) defined in~\cite{qiao2022isolation}, and crosstalk cancellation (XTC). Metrics are defined frequency-wise based on reproduced pressures measured at the left and right ears. In this work, we consider a two-listener scenario, i.e., $K=2$.

\paragraph*{IZI and IPI.}
Following~\cite{qiao2022isolation}, we activate each program individually.
Let $P^{(j)}_{k,e}(\omega)$ denote the complex reproduced pressure at listener $k$ and ear $e$ when only program $j$ is played with unit energy.
For each frequency $\omega$, we define
\begin{equation}
E^{\mathrm{tar}}_{k}(\omega)
=
\sum_{e\in\{L,R\}}\left|P^{(k)}_{k,e}(\omega)\right|^2,
\end{equation}
\begin{equation}
E^{\mathrm{int}}_{k}(\omega)
=
\sum_{e\in\{L,R\}}\sum_{j\neq k}\left|P^{(j)}_{k,e}(\omega)\right|^2,
\end{equation}
\begin{equation}
E^{\mathrm{leak}}_{k}(\omega)
=
\sum_{i\neq k}\sum_{e\in\{L,R\}}\left|P^{(k)}_{i,e}(\omega)\right|^2.
\end{equation}
The frequency-dependent IZI and IPI are
\begin{equation}
\mathrm{IZI}_k(\omega)
=
10\log_{10}\!\left(
\frac{E^{\mathrm{tar}}_{k}(\omega)}
{E^{\mathrm{leak}}_{k}(\omega)+\epsilon}
\right).
\label{eq:izi_freq}
\end{equation}

\begin{equation}
\mathrm{IPI}_k(\omega)
=
10\log_{10}\!\left(
\frac{E^{\mathrm{tar}}_{k}(\omega)}
{E^{\mathrm{int}}_{k}(\omega)+\epsilon}
\right),
\label{eq:ipi_freq}
\end{equation}
where $\epsilon$ is a small regularization constant ensuring numerical stability.

\paragraph*{Crosstalk cancellation (XTC).}
For each listener $k$, we estimate a $2\times2$ end-to-end transfer matrix $\mathbf{T}_k(\omega)$ mapping intended stereo input channels to reproduced ear signals.
The frequency-dependent XTC is defined as
\begin{equation}
\mathrm{XTC}_k(\omega)
=
10\log_{10}\!\left(
\frac{|T_{k,L,L}(\omega)|^2+|T_{k,R,R}(\omega)|^2}
{|T_{k,L,R}(\omega)|^2+|T_{k,R,L}(\omega)|^2+\epsilon}
\right).
\end{equation}

Broadband results are obtained by log-mean averaging of the dB-valued metrics over the frequency range of interest.

\section{Experimental Setup and Evaluation}
\label{sec:exp_setup}

To quantify the impact of incorporating physically informed components, we validate the four cumulative configurations (C0--C3) through real-world experiments. This section details the physical testbed and the measurement-based evaluation protocol.
All evaluations were conducted in a listening room using two static Brüel \& Kjær head-and-torso simulators (HATS).
Crucially, to ensure a fair comparison, the BSANN model architecture, training hyperparameters, and evaluation procedure remained invariant across all experiments; the sole variable was the synthesis strategy used to generate the training data, as defined in Sec.~\ref{sec:method_ablation}.

\subsection{Physical Setup}
\label{sec:exp_physical}

\begin{figure}[t]
\begin{center}
\includegraphics[width=\linewidth]{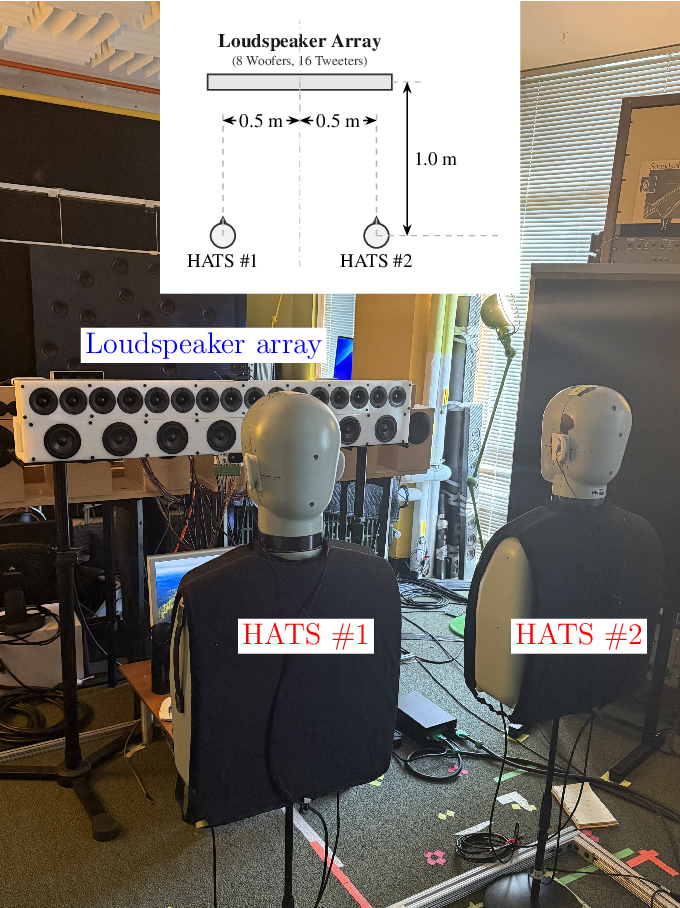}
\caption{Geometric configuration of the experimental testbed. The playback system features a 24-element loudspeaker array mounted on a rigid baffle, comprising two rows: 8 woofers ($100$--$2000$\,Hz) and 16 tweeters ($2$--$20$\,kHz). Two HATS are positioned symmetrically at a distance of 1.0\,m from the array with lateral offsets of $\pm 0.5$\,m relative to the centerline, oriented perpendicular to the array midpoint.}
\label{fig:setup}
\end{center}
\end{figure}

The loudspeaker system consists of a $L=24$-element linear array mounted on a rigid baffle at listener height. The array comprises two horizontal rows: 8 woofers operating at approximately 100--2000~Hz and 16 tweeters covering 2--20~kHz.
Two HATS are positioned symmetrically about the array centerline, located 0.5~m to the left and right of the center, respectively, at a distance of 1.0~m from the array.
The HATS are aligned vertically with the loudspeakers, with their interaural axes oriented perpendicular to the array's midpoint.
The detailed geometric configuration of the testbed is illustrated in Fig.~\ref{fig:setup}.

\subsection{Training Conditions}
\label{sec:exp_train}

We train four BSANN models corresponding to the cumulative ablation configurations C0--C3 (Table~\ref{tab:ablation}). For each model, training ATFs are synthesized with the same simulation backbone, while varying only the ATF construction pipeline: baseline point-source responses (C0), plus measured loudspeaker frequency responses of the particular loudspeakers (C1), plus analytic piston directivity on the direct path (C2), plus rigid-sphere HRTF on the direct path (C3). All network architecture choices, loss weights, optimization settings, and data splits are kept identical across C0--C3.

\subsection{Measurement-Based Evaluation Protocol}
\label{sec:exp_eval}

All configurations were evaluated under static head poses. 
For each test case, the trained BSANN generated the specific loudspeaker control filters for the fixed listener geometry. 
To validate the performance, broadband exponential sine sweep (ESS) excitation signals were processed by these predicted filters and reproduced over the loudspeaker array. 
The resulting binaural signals were captured using free-field-equalized BACCH-BM Pro in-ear microphones positioned at the entrance of the HATS' ear canals. 
The recorded sweeps were used to compute the IZI, IPI, and XTC metrics defined in Sec.~\ref{sec:method_metrics}. Unless otherwise stated, results are reported as log-frequency--weighted averages over 100--20{,}000~Hz.

\section{Results and Analysis}
\label{sec:results}

The four cumulative training configurations (C0--C3), as defined in Sec.~\ref{sec:method_ablation}, are evaluated via in-situ measurements at a single static pose under a fixed evaluation geometry. Performance metrics are computed frequency-wise (Sec.~\ref{sec:method_metrics}) and aggregated as broadband log-mean values over the $100$--$20{,}000$~Hz bandwidth (calculated as the arithmetic mean of decibel values across log-spaced frequency bins).

\subsection{Overall measured performance}
\label{sec:results_overall}

\begin{figure*}[t]
\centering

\begin{minipage}[t]{0.32\textwidth}
\centering
\includegraphics[width=\linewidth]{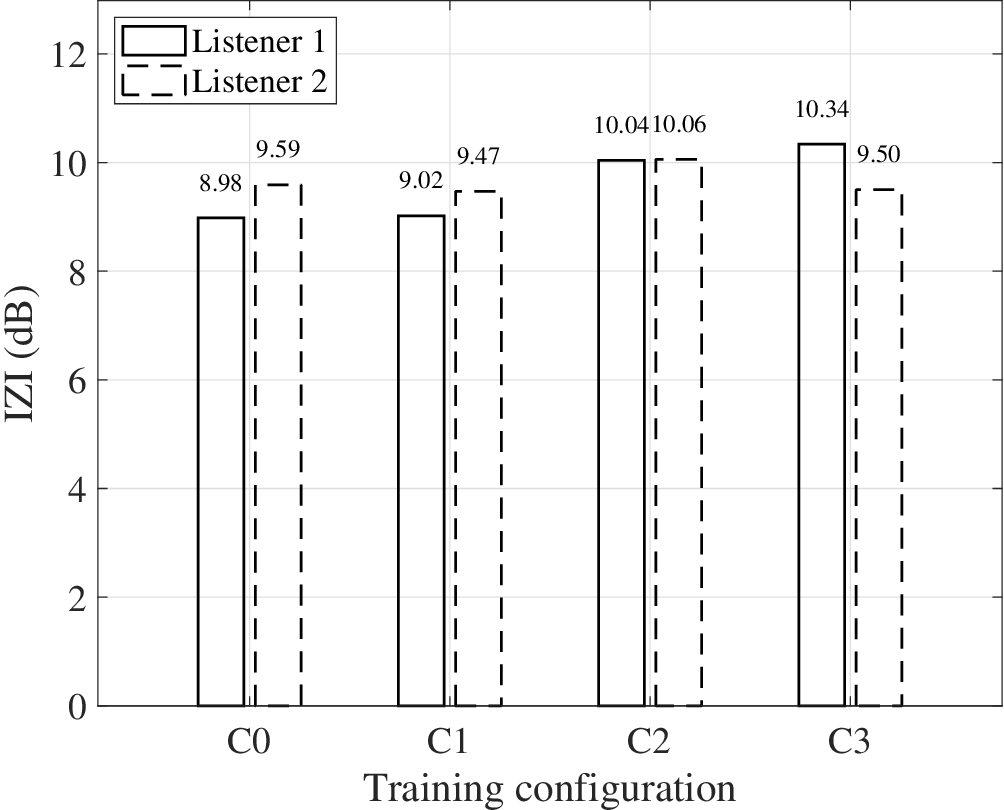}\\
{\small (a) Broadband IZI}
\end{minipage}\hfill
\begin{minipage}[t]{0.32\textwidth}
\centering
\includegraphics[width=\linewidth]{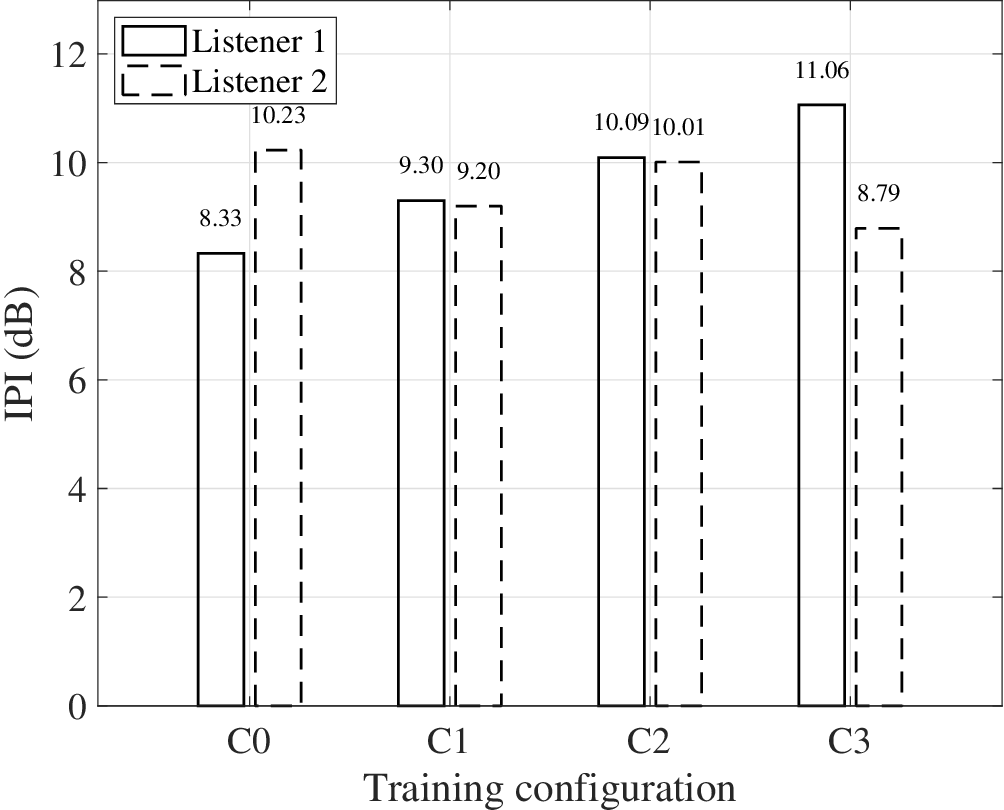}\\
{\small (b) Broadband IPI}
\end{minipage}\hfill
\begin{minipage}[t]{0.32\textwidth}
\centering
\includegraphics[width=\linewidth]{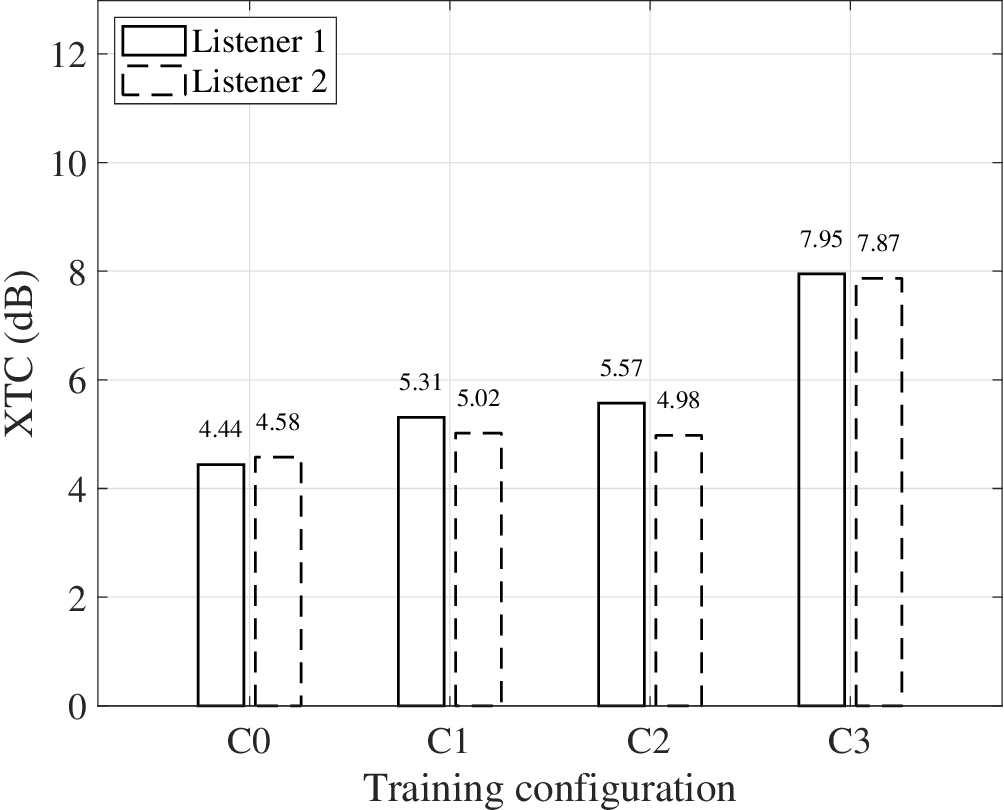}\\
{\small (c) Broadband XTC}
\end{minipage}

\vspace{0.6em}

\begin{minipage}[t]{0.32\textwidth}
\centering
\includegraphics[width=\linewidth]{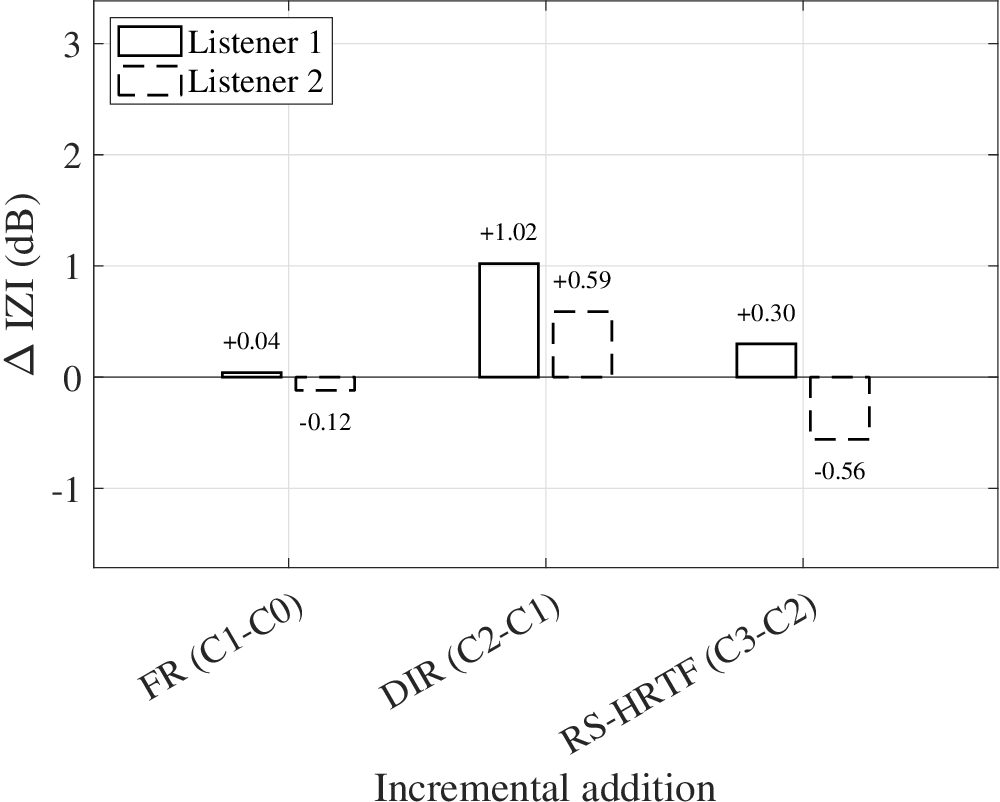}\\
{\small (d) Incremental $\Delta$IZI}
\end{minipage}\hfill
\begin{minipage}[t]{0.32\textwidth}
\centering
\includegraphics[width=\linewidth]{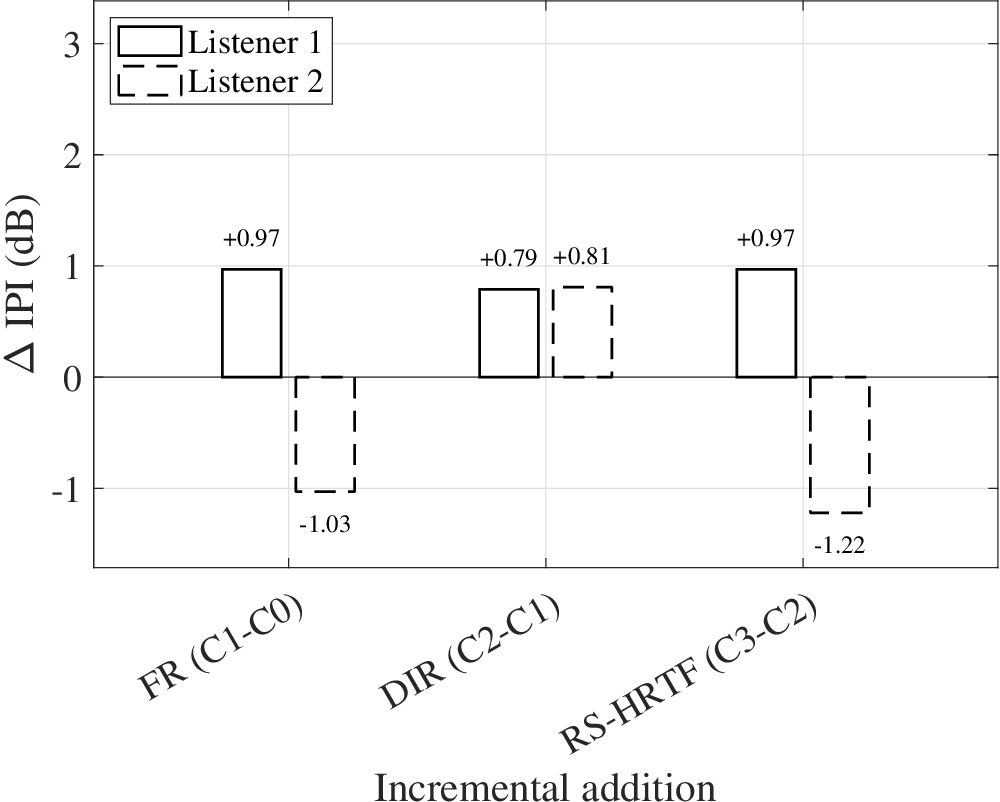}\\
{\small (e) Incremental $\Delta$IPI}
\end{minipage}\hfill
\begin{minipage}[t]{0.32\textwidth}
\centering
\includegraphics[width=\linewidth]{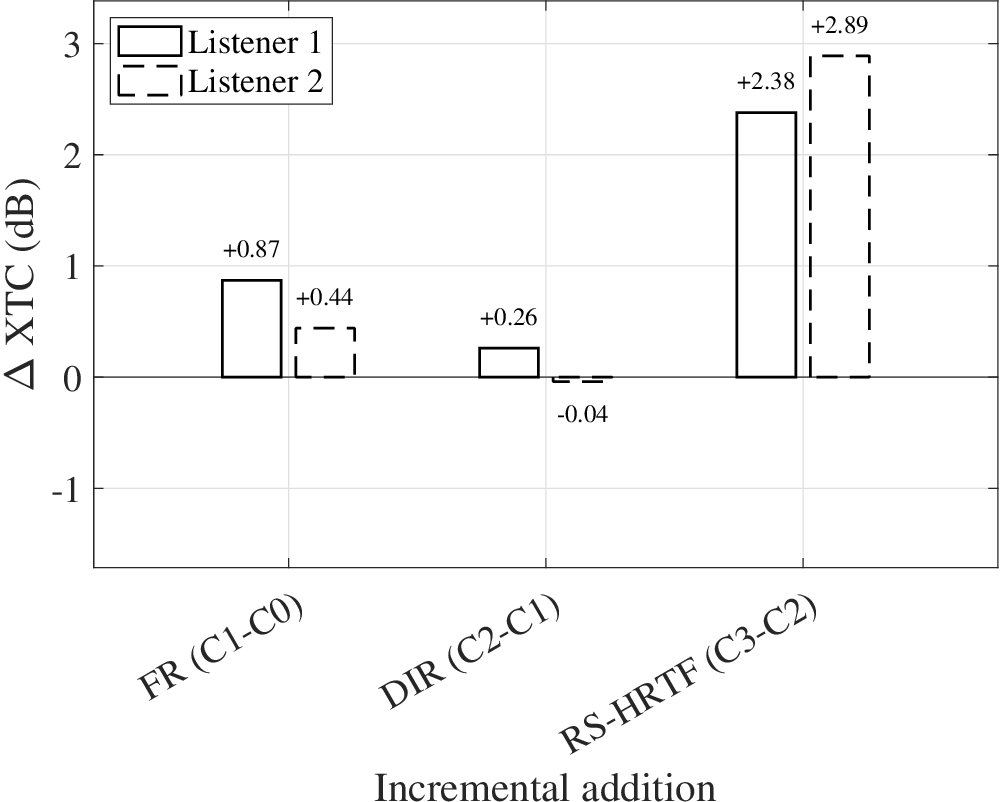}\\
{\small (f) Incremental $\Delta$XTC}
\end{minipage}

\caption{Broadband performance summary for the four cumulative configurations. Top row: log-mean metric values over 100--20{,}000~Hz for Listener~1 and Listener~2 under C0--C3. Bottom row: incremental changes under the cumulative ablation protocol (FR: C1$-$C0; DIR: C2$-$C1; RS-HRTF: C3$-$C2).}
\label{fig:bars_abs_delta}
\end{figure*}

\begin{figure*}[t]
\centering
\begin{minipage}[t]{0.32\textwidth}
\centering
\includegraphics[width=\linewidth]{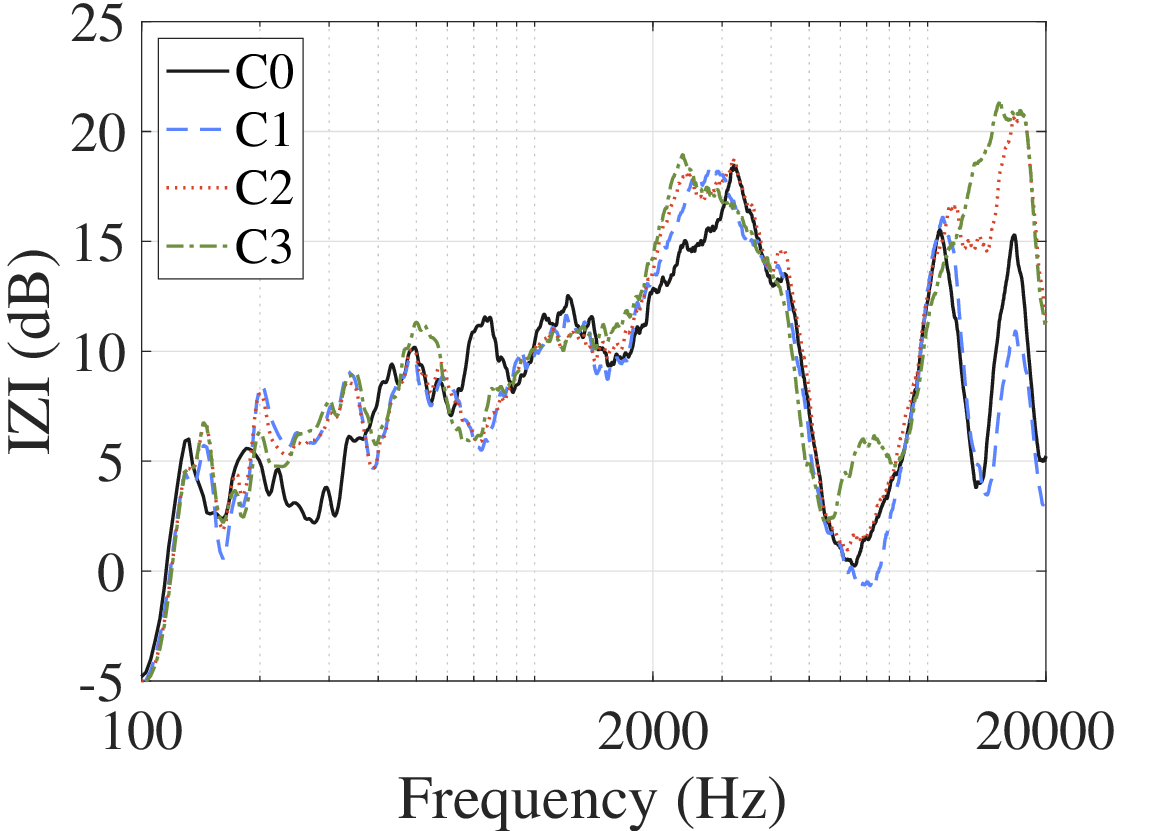}\\
{\small (a) Listener 1: IZI}
\end{minipage}\hfill
\begin{minipage}[t]{0.32\textwidth}
\centering
\includegraphics[width=\linewidth]{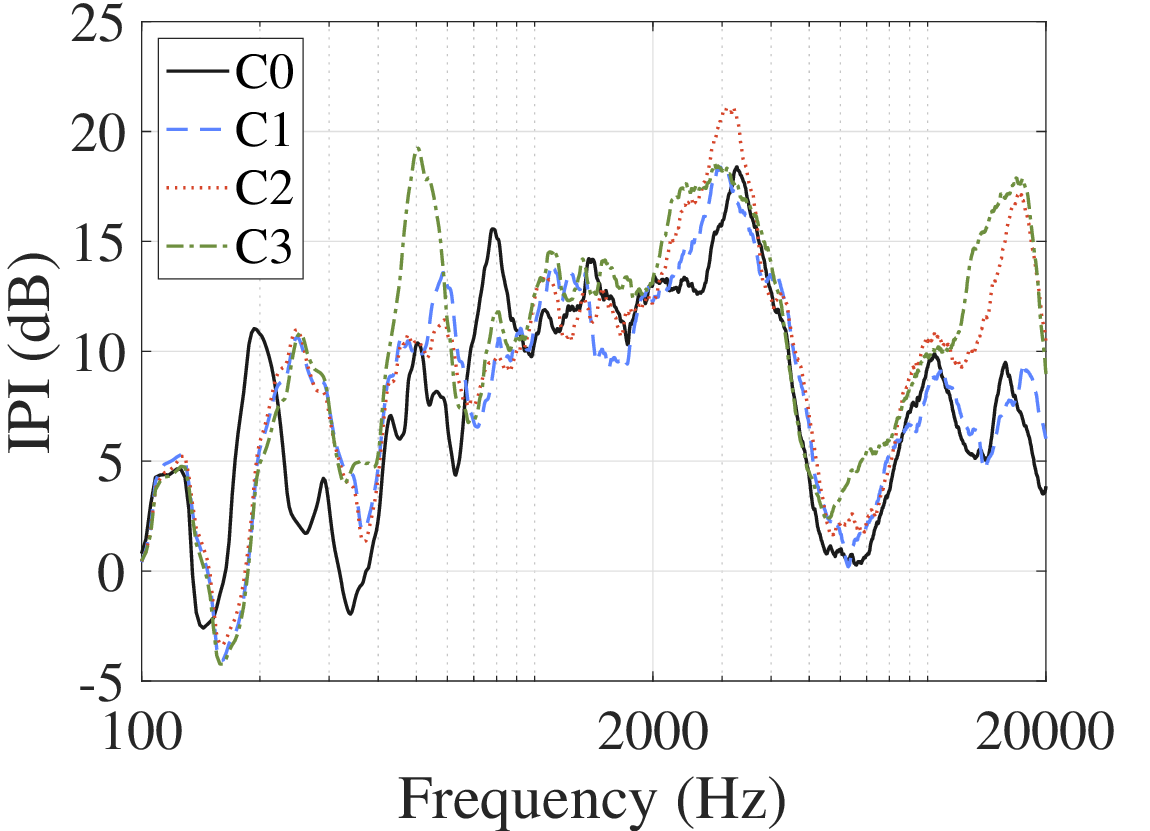}\\
{\small (b) Listener 1: IPI}
\end{minipage}\hfill
\begin{minipage}[t]{0.32\textwidth}
\centering
\includegraphics[width=\linewidth]{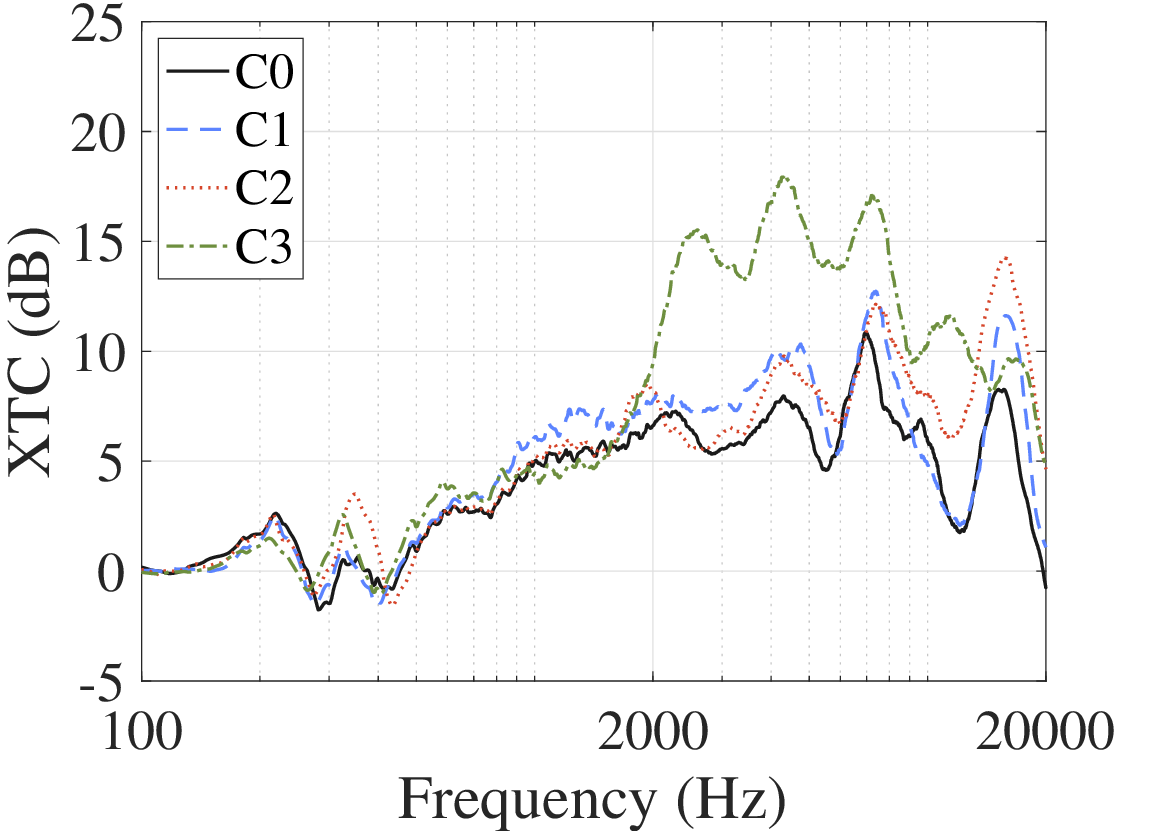}\\
{\small (c) Listener 1: XTC}
\end{minipage}

\vspace{0.6em}

\begin{minipage}[t]{0.32\textwidth}
\centering
\includegraphics[width=\linewidth]{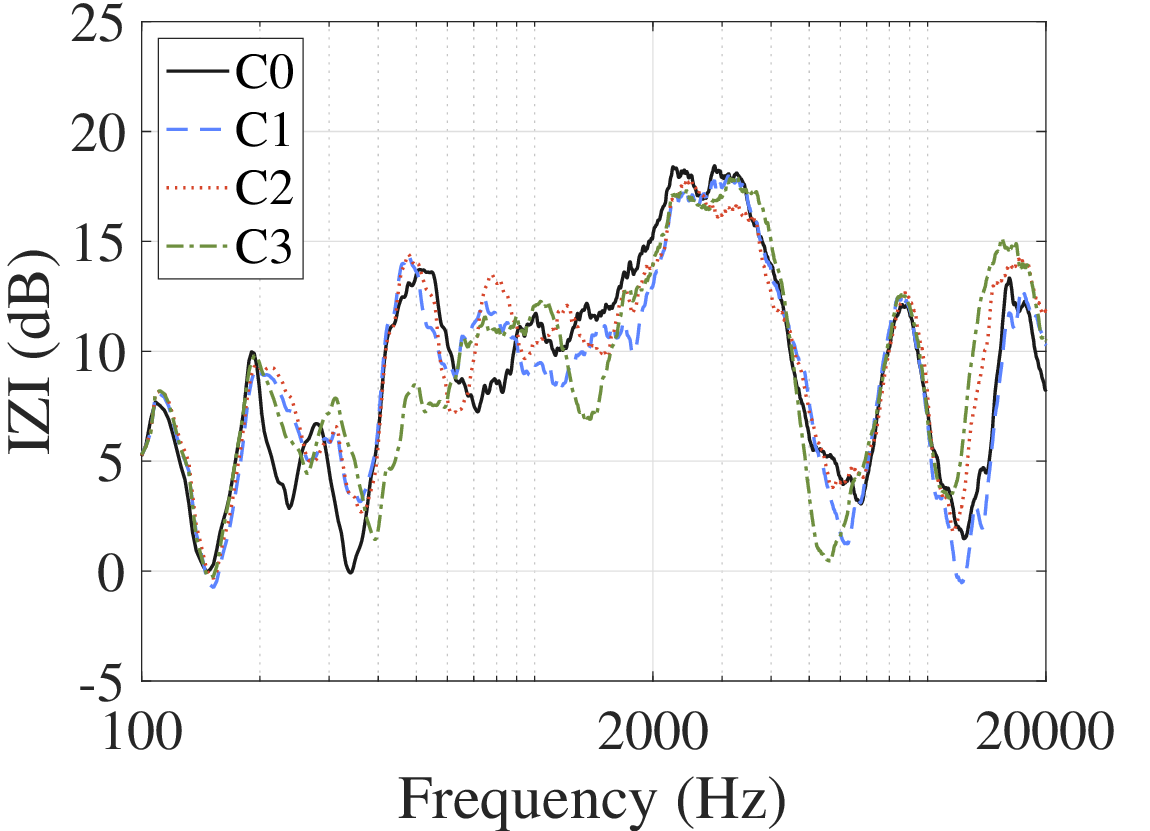}\\
{\small (d) Listener 2: IZI}
\end{minipage}\hfill
\begin{minipage}[t]{0.32\textwidth}
\centering
\includegraphics[width=\linewidth]{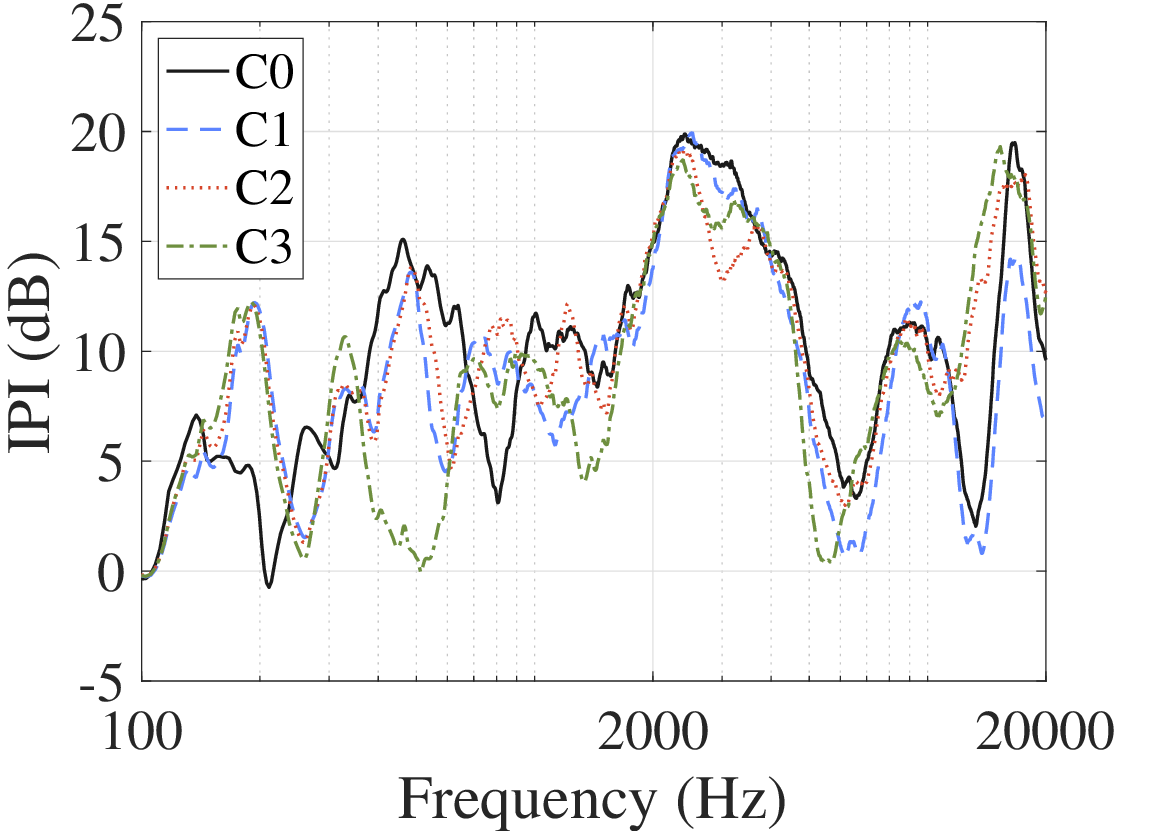}\\
{\small (e) Listener 2: IPI}
\end{minipage}\hfill
\begin{minipage}[t]{0.32\textwidth}
\centering
\includegraphics[width=\linewidth]{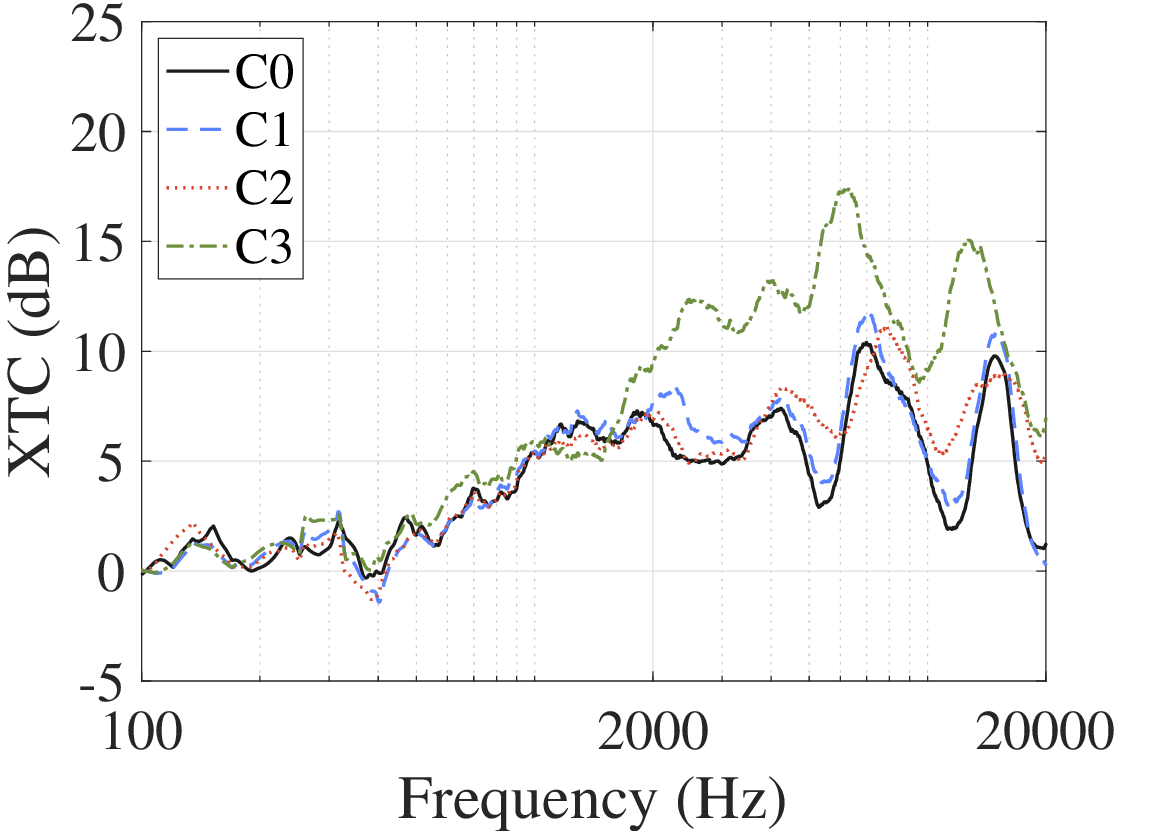}\\
{\small (f) Listener 2: XTC}
\end{minipage}

\caption{Frequency-dependent IZI, IPI, and XTC for the four cumulative training configurations, evaluated at a single static pose. Legends correspond to C0 (baseline point-source simulation), C1 (+FR), C2 (+FR+DIR), and C3 (+FR+DIR+RS-HRTF). Top row: Listener~1. Bottom row: Listener~2.}
\label{fig:freq_metrics}
\end{figure*}

Fig.~\ref{fig:bars_abs_delta}(a--c) summarizes the broadband log-mean metrics for each listener, while Fig.~\ref{fig:freq_metrics} depicts the corresponding frequency-dependent characteristics. Analysis of the measured data reveals three key observations.

First, XTC is substantially enhanced by incorporating head scattering (RS-HRTF). Configuration C3 yields the highest XTC for both listeners (XTC$_1$: $7.95$~dB; XTC$_2$: $7.87$~dB), raising the listener-averaged XTC from $4.51$~dB in the baseline (C0) to $7.91$~dB in C3 [Fig.~\ref{fig:bars_abs_delta}(c)]. 

This trend is clearly reflected in the XTC spectra in Fig.~\ref{fig:freq_metrics}, where C3 separates markedly from the other configurations in the mid-to-high-frequency range (starting around 2~kHz). Conversely, below 1~kHz, performance across configurations remains broadly comparable, indicating that low-frequency binaural separation is less sensitive to the inclusion of head scattering in this setup. Notably, C1 and C2 provide only modest XTC changes relative to C0, whereas the dominant gain occurs when enabling RS-HRTF in C3.

Second, sound-zone separation metrics (IZI and IPI) are primarily driven by source directivity modeling. The optimal listener-averaged IZI and IPI are achieved by configuration C2 (+FR+DIR), reaching $10.05$~dB for both metrics [Fig.~\ref{fig:bars_abs_delta}(a,b)]. This observation aligns with Fig.~\ref{fig:freq_metrics}, where C2 demonstrates a sustained uplift in the mid-frequency region, corresponding to the bandwidth where energy steering and leakage/interference suppression are most effective.

Finally, Fig.~\ref{fig:bars_abs_delta}(a,b) reveals an inter-listener asymmetry in sound-zone separation. For Listener~1, IZI and IPI improve monotonically with increasing physical fidelity, peaking at C3 (IZI$_1$: $8.98 \to 10.34$~dB; IPI$_1$: $8.33 \to 11.06$~dB). In contrast, for Listener~2, sound-zone separation improves from C0 to C2, with C2 yielding the best overall performance (IZI$_2$: $9.59 \to 10.06$~dB; IPI$_2$: $10.23 \to 10.01$~dB), while a further increase in physical fidelity to C3 leads to a degradation in sound-zone separation, most notably in IPI. This non-monotonic trend is further supported by the frequency-dependent curves, where the C3 response falls below that of C2 in specific sub-bands.

\subsection{Incremental contribution of FR, DIR, and RS-HRTF}
\label{sec:results_incremental}

To isolate the incremental effect of each physical component under the cumulative protocol, Fig.~\ref{fig:bars_abs_delta}(d--f) reports broadband log-mean changes (100--20{,}000~Hz) separately for Listener~1 and Listener~2, corresponding to the successive additions FR (C1$-$C0), DIR (C2$-$C1), and RS-HRTF (C3$-$C2).

\paragraph*{FR (C1 vs.\ C0).}
Adding anechoically measured loudspeaker frequency responses of the particular loudspeakers yields a modest but consistent improvement in XTC for both listeners ($\Delta$XTC$_1{=}{+}0.87$~dB, $\Delta$XTC$_2{=}{+}0.44$~dB; Fig.~\ref{fig:bars_abs_delta}(f)). In Fig.~\ref{fig:freq_metrics}, this appears as moderate curve shifts relative to C0 across portions of the band. For sound-zone separation metrics, FR has a minor effect on IZI but substantially reduces the inter-listener asymmetry in IPI. While C0 exhibits a large disparity (IPI$_2$ exceeds IPI$_1$ by 1.90~dB), adding FR increases IPI$_1$ by 0.97~dB and decreases IPI$_2$ by 1.03~dB, resulting in nearly balanced performance (IPI$_1{=}9.30$~dB, IPI$_2{=}9.20$~dB) [Fig.~\ref{fig:bars_abs_delta}(b)].

This convergence suggests that incorporating FR reduces loudspeaker-induced spectral mismatch and redistributes the interference-suppression trade-off across the two listeners under the fixed objective.

\paragraph*{DIR (C2 vs.\ C1).}
Introducing analytic piston directivity produces the most consistent gains in sound-zone separation across both listeners: IZI improves by $\Delta$IZI$_1{=}{+}1.02$~dB and $\Delta$IZI$_2{=}{+}0.59$~dB, while IPI improves by $\Delta$IPI$_1{=}{+}0.79$~dB and $\Delta$IPI$_2{=}{+}0.81$~dB (Fig.~\ref{fig:bars_abs_delta}(d,e)). In Fig.~\ref{fig:freq_metrics}, this corresponds to a broadband uplift of the separation curves, consistent with more realistic frequency-dependent energy steering and reduced leakage/interference once non-omnidirectional radiation is modeled. 

Changes in XTC at this stage are small and listener-dependent ($\Delta$XTC$_1{=}{+}0.26$~dB, $\Delta$XTC$_2{=}{-}0.04$~dB; Fig.~\ref{fig:bars_abs_delta}(f)), indicating that DIR primarily benefits sound-zone separation (IZI/IPI) via spatial energy confinement, while having limited impact on binaural channel decoupling in this setup.

\paragraph*{RS-HRTF (C3 vs.\ C2).}
Adding rigid-sphere head scattering dominates the improvement in binaural separation, boosting XTC by $\Delta$XTC$_1{=}{+}2.38$~dB and $\Delta$XTC$_2{=}{+}2.89$~dB (Fig.~\ref{fig:bars_abs_delta}(f)). In Fig.~\ref{fig:freq_metrics}, the improvement is most evident from the mid-frequency region upward (starting around 2~kHz), where head-induced binaural differences are strong. At the same time, the effect on sound-zone separation metrics becomes strongly listener-dependent: for Listener~1, both zone isolation and inter-program interference improve ($\Delta$IZI$_1{=}{+}0.30$~dB, $\Delta$IPI$_1{=}{+}0.97$~dB), whereas for Listener~2 they degrade ($\Delta$IZI$_2{=}{-}0.56$~dB, $\Delta$IPI$_2{=}{-}1.22$~dB) [Fig.~\ref{fig:bars_abs_delta}(d,e)]. 

The substantial XTC gain suggests that incorporating head-induced scattering and shadowing is critical for mid-to-high-frequency binaural separation under measured playback. The divergent behavior in IZI and IPI indicates a mild, geometry-dependent trade-off: modeling the local binaural scattering structure that benefits XTC can coincide with changes in energy steering and leakage/interference suppression, leading to listener-dependent shifts in both IZI and IPI.

\section{Conclusion}
\label{sec:conclusion}

This paper presented a controlled, cumulative ablation study to quantify how physically informed acoustic components in the ATF-generation pipeline affect neural personal sound zone rendering. Using the BSANN framework~\cite{Jiang2026BSANN} with all network and training settings held fixed, we evaluated four training configurations (C0--C3) on in-situ measurements with two HATS at a single static pose. Performance was assessed using frequency-dependent IZI, IPI, and XTC, with broadband summaries reported as log-mean values over 100--20{,}000~Hz.

The results provide clear component-wise attribution. Adding measured loudspeaker frequency responses of the particular loudspeakers (FR) produced a small but consistent XTC improvement for both listeners and, importantly, reduced inter-listener imbalance in IPI, indicating that accurate transducer spectral characterization primarily serves as a practical calibration step, reducing spectral mismatch and inter-listener imbalance even when improvements in peak isolation are modest. Introducing analytic piston directivity (DIR) yielded the most consistent gains in sound-zone separation (IZI and IPI) across both listeners, while having negligible impact on XTC, suggesting that directivity primarily improves spatial energy confinement and leakage/interference suppression rather than binaural channel decoupling. Incorporating rigid-sphere head scattering (RS-HRTF) dominated the improvement in binaural separation, producing the largest XTC gains, particularly above 2~kHz. At the same time, RS-HRTF inclusion led to listener-dependent shifts in IZI/IPI, indicating a mild, geometry-dependent trade-off between maximizing XTC and maintaining sound-zone separation under the fixed renderer and objective.

From a practical perspective, these findings suggest a prioritization strategy when measurement or implementation resources are limited:
\begin{itemize}
\item for applications primarily targeting sound-zone separation (IZI/IPI), modeling loudspeaker directivity is the most impactful physical component;
\item for binaural reproduction quality and crosstalk cancellation (XTC), incorporating a head-scattering model is essential; and
\item measured loudspeaker spectral responses are beneficial for reducing sim-to-real spectral mismatch and balancing performance across listeners.
\end{itemize}

Future work will extend this analysis beyond a single static pose to dynamic and multi-position evaluations, and will investigate replacing the rigid-sphere HRTF with individualized or groupwise HRTFs drawn from large databases to further enhance binaural accuracy, especially at higher frequencies where pinna and torso effects dominate, all under the same cumulative ablation framework.

\bibliographystyle{unsrtnat} 
\bibliography{refs}

\end{document}